# Quantum plasmonic sensing by Hong-Ou-Mandel interferometry

**SEUNGJIN YOON,**[1,2,3,*] **YU SUNG CHOI,**[3] **MARK TAME,**[4] **JAE WOONG YOON,**[3] **SERGEY V. POLYAKOV,**[1,5] **AND CHANGHYOUP LEE**[6,†]

[1]*National Institute of Standards and Technology, Gaithersburg, MD 20899, USA*
[2]*Joint Quantum Institute, University of Maryland, College Park, MD 20742, USA*
[3]*Department of Physics, Hanyang University, Seoul 04763, Republic of Korea*
[4]*Department of Physics, Stellenbosch University, Matieland 7602, South Africa*
[5]*Department of Physics, University of Maryland, College Park, MD 20742, USA*
[6]*Korea Research Institute of Standards and Science, Daejeon 34113, Republic of Korea*
[*]*seungjin.yoon@nist.gov*
[†]*changhyoup.lee@gmail.com*

**Abstract:** We propose a quantum plasmonic sensor using Hong-Ou-Mandel (HOM) interferometry that measures the refractive index of an analyte, embedded in a plasmonic beam splitter composed of a dual-Kretschmann configuration, which serves as a frustrated total internal reflection beamsplitter. The sensing performance of the HOM interferometry, combined with single-photon detectors, is evaluated through Fisher information for estimation of the refractive index of the analyte. This is subsequently compared with the classical benchmark that considers the injection of a coherent state of light into the plasmonic beamsplitter. By varying the wavelength of the single photons and the refractive index of the analyte, we identify a wide range where a 50 % quantum enhancement is achieved and discuss the observed behaviors in comparison with the classical benchmark. We expect this study to provide a useful insight into the advancement of quantum-enhanced sensing technologies, with direct implications for a wide range of nanophotonic beamsplitter structures.

## 1. Introduction

Surface plasmons polaritons (SPPs) are electromagnetic fields coupled to collective oscillations of electrons at the interface between a metallic and a dielectric medium [1]. They are strongly confined at the metal surface, consequently leading to a large density of electromagnetic states. Such a feature makes plasmonic structures more sensitive to a change in their environment than conventional photonic ones [2–4]. Furthermore, the small mode volume induced by a strong field confinement also enables sub-diffraction limited imaging [5, 6]. Consequently, plasmonic sensors have found widespread applications in practical photonic sensing [7, 8], and have finally achieved successful commercialization.

These conventional plasmonic sensors employ classical light, so their signal-to-noise ratio is shot-noise limited [9, 10], which originates from a Poisson photon number distribution embedded in classical light, i.e., the so-called coherent state of light [11]. With the recent advances in quantum technology, many studies have suggested to use quantum states of light in plasmonic sensing to reduce the noise below the shot-noise limit [11–13]. Such a new methodology, called 'quantum plasmonic sensing' can be considered in two categories: phase and intensity sensing, depending on which useful quantum state of light and detection scheme is used [14]. These new types of plasmonic sensors have been well framed in terms of quantum parameter estimation theory, which has suggested various quantum plasmonic sensors offering a range of quantum enhancement [4, 11–14].

In this work, we propose a new kind of quantum plasmonic sensing, inspired by recent studies that exploit the Hong-Ou-Mandel (HOM) interference for given two single photons [15–17]. The HOM interference is observed when two incident bosons are indistinguishable [18, 19].

Interestingly, it has been demonstrated that SPPs follow bosonic statistics, leading to the observation of a HOM dip in experiments [20–25] and even anti-coalescence interference that is induced by photon loss in a metal or a coherent perfect absorber [21, 26, 27]. In addition, precise measurement of the refractive index is important for detecting slight changes in various sensing systems, such as biosensors, gas detection, and in semiconductor inspection [28–30]. Plasmonic sensors are commonly used to accurately measure changes in the refractive index, e.g. to estimate the concentration of a solute in an aqueous solution [9, 11]. This motivates us to study the possibility of using plasmonic HOM interferometry to measure the refractive index of an analyte.

The work is organized as follows: In Section 2, we introduce interferometry with the dual Kretschmann plasmonic beamsplitter (BS) we propose and the optical response using a theoretical description we develop. In Section 3, we evaluate the sensing performance of the plasmonic HOM sensor in comparison with the classical benchmark. We also discuss the quantum enhancement in the estimation of the refractive index parameter of an analyte. In Section 4, we summarize our study and provide an outlook on future work.

## 2. Interferometry with a plasmonic beam splitter

### 2.1. Plasmonic beam splitter

When two single photons simultaneously arrive at a BS from different input ports, the coincidence probability (or count) between the two output ports depends on the overlap of the two single-photon wavepackets. If the wavepackets are fully overlapped, i.e., the two single photons are completely indistinguishable, the coincidence probability drops to zero, known as a HOM dip, indicating that the two photons come out only either of the two output ports together, or 'bunched'. On the other hand, if they are completely distinguishable, the coincidence probability becomes 0.5. Thus, the coincidence probability varies between 0 and 0.5, providing a measure of the indistinguishability of the two incident single photons. This can be exploited to infer an associated parameter under investigation, e.g., a temporal interval between two identical single-photon wavepackets [15]. This is the basic idea of HOM based-sensing and has been exploited for various applications [16, 17].

In addition to the characteristics of single-photon wavepackets, the coincidence probability also depends on the features of the BS, e.g., transmission/reflection coefficients and their phases. This implies that one can also infer parameters that would affect the properties of a BS from the HOM signals, for identical two single-photon wavepackets, which applies to the scheme we propose in this work. In this case, if the transmission/reflection coefficients vary the coincidence probability can take a value between 0 and 1, contrary to the case of a common BS, where the reflectance and transmittance are equal, i.e. $R = T = 0.5$.

Let us consider a HOM interferometry setup with a plasmonic BS that consists of two symmetric prisms coated with gold metal films and the sample layer with a refractive index $n_s$, as shown in Fig. 1. The plasmonic BS we consider serves as a frustrated total internal reflection BS [31], but with a metallic film on each prism to exploit the excitation of SPPs [32]. When light is incident on the prism at an incidence angle greater than the critical angle, SPPs are excited on the metallic surface on the sample side, as in the Kretschmann configuration [1]. The generated SPPs may propagate along the metallic surface, but their evanescent field component perpendicular to the surface excites SPPs on the opposite metallic film of the second Kretschmann configuration in the layered structure. Subsequently, the secondary generated SPPs can be coupled out through the prism, which is the transmission channel of the plasmonic BS we propose. The transmittance spectrum $T(\omega)$ depends on both geometric and material properties of the plasmonic BS structure, which also determine the reflectance spectrum $R(\omega)$. Unlike a conventional BS, however, the plasmonic BS involves inevitable absorption $A(\omega)$ due to the Ohmic loss in a metal, i.e., $T(\omega) + R(\omega) + A(\omega) = 1$. This mechanism applies to both the input

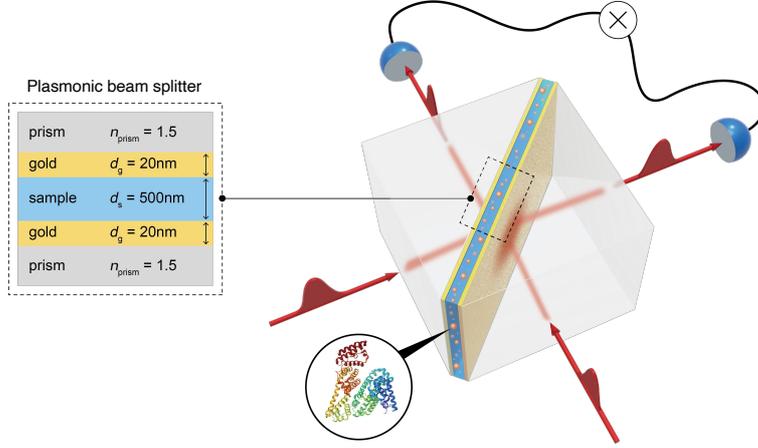

Fig. 1. A schematic of a plasmonic Hong-Ou-Mandel (HOM) interferometer. Two identical single photons are incident to the plasmonic BS composed of a dual-Kretschmann configuration, inside of which a sample layer is embedded. At the output ports, a coincidence detection is made with single-photon detectors. (Inset top left) A top-view of the plasmonic BS that consists of five layers. (Inset bottom left) A protein that makes up the sample layer with a refractive index $n_s$ that depends on the concentration.

modes of the prism shown in Fig. 1.

### 2.2. Hong-Ou-Mandel interference

The quantum input-output relation for a lossy BS can be generally written as

$$\begin{pmatrix} \hat{b}_1(\omega) \\ \hat{b}_2(\omega) \end{pmatrix} = \begin{pmatrix} t_{12}(\omega) & r_{12}(\omega) \\ r_{21}(\omega) & t_{21}(\omega) \end{pmatrix} \begin{pmatrix} \hat{a}_1(\omega) \\ \hat{a}_2(\omega) \end{pmatrix} + \begin{pmatrix} \hat{f}_1(\omega) \\ \hat{f}_2(\omega) \end{pmatrix}, \quad (1)$$

where $\hat{a}_{1(2)}(\omega)$, $\hat{b}_{1(2)}(\omega)$ and $\hat{f}_{1(2)}(\omega)$ denote the annihilation operators of the input, output, and environment (Ohmic loss) mode, respectively. These annihilation operators satisfy the commutation relations [33], written as

$$[\hat{a}_j(\omega), \hat{a}_k(\omega')] = 0 = [\hat{b}_j(\omega), \hat{b}_k(\omega')], \quad (2)$$

$$[\hat{a}_j(\omega), \hat{a}_k^\dagger(\omega')] = \delta_{jk}\delta(\omega - \omega') = [\hat{b}_j(\omega), \hat{b}_k^\dagger(\omega')], \quad (3)$$

where $j, k \in \{1, 2\}$. Here, the transmission and reflection coefficients are simplified to $t_{12}(\omega) = t_{21}(\omega) \equiv |t(\omega)|e^{i\phi_t(\omega)}$ and $r_{12}(\omega) = r_{21}(\omega) \equiv |r(\omega)|e^{i\phi_r(\omega)}$, respectively. This simplification is enabled due to the symmetric nature of the plasmonic BS considered in our work, both in terms of its geometry and material properties. It is also important to note that the phases $\phi_t(\omega)$ and $\phi_r(\omega)$ can assume arbitrary values in our case due to the presence of loss [34]. This contrasts with the lossless case, where the phase condition $|\phi_t(\omega) - \phi_r(\omega)| = \pi/2$ modulo $\pi$ must be satisfied for a symmetric BS [35]. To determine the transmission and reflection coefficients of the plasmonic BS in this work, we employ the transfer matrix method, suitable for describing the propagation of electromagnetic waves through multilayer structures [36, 37]. The detailed calculation is provided in Appendix A.

Consider two single-photon wavepackets incident on the two input ports and that an individual single-photon state wavepacket can be generally written in the frequency domain as

$$|1_j\rangle = \int_0^\infty d\omega_j\, \xi(\omega_j) \hat{a}_j^\dagger(\omega_j)|0_j\rangle, \tag{4}$$

where $\xi(\omega_j)$ is a spectral profile satisfying $\int |\xi(\omega_j)|^2 d\omega = 1$. Applying the input-output relation of Eq. (1) to the input state $|\psi_{\text{in}}\rangle = |1_1\rangle|1_2\rangle$, one can find the output state $|\psi_{\text{out}}\rangle$. For a coincidence counting scheme, we consider two single-photon detectors, which have been widely used in most HOM interferometers [15–17]. They lead to the three measurement outcomes: Neither detector clicks, either detector clicks, and both detectors click. The probabilities for the number of clicks can be written in terms of those for photon numbers in the two output ports as

$$P(0_{\text{click}}|n_s) = P(0_1, 0_2|n_s), \tag{5}$$
$$P(1_{\text{click}}|n_s) = P(1_1, 0_2|n_s) + P(2_1, 0_2|n_s) + P(0_1, 1_2|n_s) + P(0_1, 2_2|n_s), \tag{6}$$
$$P(2_{\text{click}}|n_s) = P(1_1, 1_2|n_s), \tag{7}$$

where $P(l_{\text{click}}|n_s)$ is the probability of $l$ detectors producing a click conditioned on the refractive index $n_s$ in the HOM interferometry, and $P(l_1, l_2|n_s)$ denotes the probability of photon numbers $l_1$ and $l_2$ present in the two output ports, respectively, conditioned on the refractive index $n_s$. A similar mapping has been implemented in a previous study with a lossless model [15], but note that here we take into account photon loss. In what follows in this section, we drop the refractive index $n_s$ in the conditional probabilities to simplify the presentation, but reinstate it in the next section for clarity. The above photon number probabilities can be easily obtained using the Kelley-Kleiner counting formulae, assuming an ideal detection, e.g., unit-detection efficiency and infinite counting time [33, 34, 38]. This allows us to use the average value of the number operator in the $j$th output port, defined as

$$\hat{N}_j = \int_0^\infty d\omega\, \hat{b}_j^\dagger(\omega) \hat{b}_j(\omega). \tag{8}$$

Due to the considered plasmonic BS being symmetric, as described above, the six probabilities for photon numbers in the output ports reduce to the four probabilities given as [33]

$$P(1_1, 1_2) = \langle \hat{N}_1 \hat{N}_2 \rangle, \tag{9}$$
$$P(2_1, 0_2) = \frac{1}{2}\langle \hat{N}_1(\hat{N}_1 - 1)\rangle = \frac{1}{2}\langle \hat{N}_2(\hat{N}_2 - 1)\rangle = P(0_1, 2_2), \tag{10}$$
$$P(1_1, 0_2) = \langle \hat{N}_1 \rangle - \frac{1}{2}\langle \hat{N}_1(\hat{N}_1 - 1)\rangle - \frac{1}{2}\langle \hat{N}_2(\hat{N}_2 - 1)\rangle - \langle \hat{N}_1 \hat{N}_2 \rangle = P(0_1, 1_2), \tag{11}$$
$$P(0_1, 0_2) = 1 - \langle \hat{N}_1 \rangle - \langle \hat{N}_2 \rangle + \langle \hat{N}_1 \hat{N}_2 \rangle + \frac{1}{2}\langle \hat{N}_1(\hat{N}_1 - 1)\rangle + \frac{1}{2}\langle \hat{N}_2(\hat{N}_2 - 1)\rangle. \tag{12}$$

The calculation of the average values (expectation values) of the number operators $\hat{N}_j$ involves the integration over frequency. The frequency-dependent Fresnel coefficients $t(\omega)$ and $r(\omega)$ are crucial in determining the transmission or reflection of the single-photon wavepackets through the plasmonic BS. As we are interested in the regime where a single-mode approximation is valid, i.e., the wavepackets are sufficiently narrow around the central frequencies $\omega_0$ with respect to the frequency-dependent behaviors of $t(\omega)$ and $r(\omega)$ (see Appendix D for the validity demonstrated by the treatment of continuum modes), we can write the average of the number operators at the central frequency $\omega_0$ as

$$\langle \hat{N}_1 \rangle = T + R = \langle \hat{N}_2 \rangle, \tag{13}$$
$$\langle \hat{N}_1(\hat{N}_1 - 1)\rangle = 4TR = \langle \hat{N}_2(\hat{N}_2 - 1)\rangle, \tag{14}$$
$$\langle \hat{N}_1 \hat{N}_2 \rangle = T^2 + R^2 + 2TR\cos 2\phi_{\text{tr}}, \tag{15}$$

where $T = |t(\omega_0)|^2$, $R = |r(\omega_0)|^2$, and $\phi_{tr} = \phi_r(\omega_0) - \phi_t(\omega_0)$ is the difference between the transmission and reflection phases. Note that the above expressions include the effect of photon loss, i.e., $T + R < 1$.

For the classical benchmark, we consider the coherent state input $|\alpha\rangle|\beta\rangle$ that is sent into the plasmonic BS under the single-mode approximation. In this case, it is reasonable to consider photon number-resolving detectors instead of single-photon detectors, which corresponds to a typical intensity detection scheme that is normally considered to set the shot-noise limit or standard quantum limit in quantum sensing and metrology [39, 40]. This leads to the probabilities of the photon numbers in the two output ports, written as

$$P^{(C)}(l_1, l_2) = e^{-\langle \hat{N}_1^{(C)} \rangle - \langle \hat{N}_2^{(C)} \rangle} \frac{\langle \hat{N}_1^{(C)} \rangle^{l_1} \langle \hat{N}_2^{(C)} \rangle^{l_2}}{l_1! l_2!}, \quad (16)$$

where $\langle \hat{N}_1^{(C)} \rangle = T|\alpha|^2 + R|\beta|^2 + 2\sqrt{TR}|\alpha||\beta|\cos(\phi_{tr} - \phi_{\alpha\beta})$ and $\langle \hat{N}_2^{(C)} \rangle = T|\alpha|^2 + R|\beta|^2 + 2\sqrt{TR}|\alpha||\beta|\cos(\phi_{tr} + \phi_{\alpha\beta})$ for $\phi_{\alpha\beta} = \arg(\alpha) - \arg(\beta)$. For a fair comparison with the HOM sensing scheme, we set $|\alpha|^2 = |\beta|^2 = 1$ in this work, but $\phi_{\alpha\beta}$ is optimized to maximize the Fisher information, as will be discussed later. So far, we have obtained the probabilities of the measurement outcomes for the two single-photon input state and the classical benchmark state with the plasmonic BS, respectively. In the next section, we investigate their sensing performance using quantum parameter estimation theory.

### 3. Results and discussion

In the interferometer with the considered plasmonic BS, whose behavior in the quantum and classical regimes was outlined in the previous section, the refractive index $n_s$ of the sample can be inferred from a measurement followed by an appropriate estimator. The estimated values of $n_s$ follow a particular distribution of an estimate over sampling and its variance is fundamentally lower-bounded even when all external noises are removed. In quantum parameter estimation theory, the lower-bound to the variance can be obtained by the Cramér-Rao inequality when a locally unbiased estimator is assumed to be used, and it is written as [41, 42]

$$\text{Var}(n_s) \geq \frac{1}{I(n_s)}, \quad (17)$$

where $I(n_s)$ denotes the Fisher information with respect to the refractive index $n_s$. In such a way, the variance of an estimate $n_s$ can be interpreted as a precision of the plasmonic sensor in the HOM interferometer influenced by the changes of the refractive index. Here, the Fisher information $I(n_s)$ is defined using a statistical model $P(m|n_s)$, which is a set of probabilities for measurement outcomes $m$ conditioned on the value of the sample parameter $n_s$, as given in Eqs. (5)–(7). The Fisher information $I(n_s)$ can be calculated as

$$I(n_s) = \sum_m \frac{1}{P(m|n_s)} \left( \frac{\partial P(m|n_s)}{\partial n_s} \right)^2. \quad (18)$$

where $P(m|n_s)$ is the probability of obtaining a measurement outcome $m$. Hence, this approach allows us to evaluate the Cramér-Rao bound in Eq. (17) for a given sensing scheme. For the purpose of this work, we calculate the Fisher information for the considered plasmonic HOM sensor and its classical counterpart, which we denote $I^{(HOM)}(n_s)$ and $I^{(C)}(n_s)$, respectively. From here, we drop off the argument $n_s$ in the Fisher information $I$ for convenience, unless it needs to be noted.

The probabilities $P(m|n_s)$ that appear in the Fisher information for the quantum and classical cases depend on the transmittance, reflectance and absorption of the plasmonic BS according to the theory outlined in the previous section. In Fig. 2(a) and (b) we show how these quantities change with the angle of incidence and refractive index $n_s$ for light at $\lambda = 800$ nm as an example. In Fig. 2(c) we also show the coincidence probability for the quantum case at a fixed incidence angle of $\theta_{in} = 70°$ past the critical angle over a range of $n_s$. Over this range, the coincidence probability reaches a minimum at $n_s \approx 1.31$ corresponding to the HOM dip for the system, where the transmittance is equal to the reflectance. The coincidence probability over the range of $n_s$ is then combined with the other probabilities to form the Fisher information.

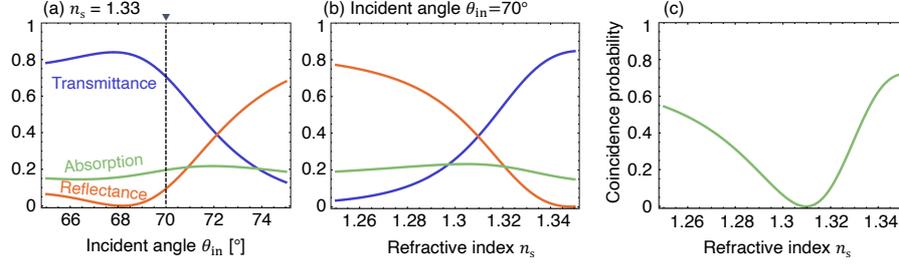

Fig. 2. (a) Transmittance, reflectance, and absorption curves of the plasmonic BS with the incidence angle $\theta_{in}$ for a sample with $n_s = 1.33$. (b) Transmittance, reflectance, and absorption curves of the plasmonic BS with the refractive index $n_s$ at a fixed incidence angle of 70° [corresponding to dashed line in (a)]. It can be shown that 70° is greater than the critical angles given by the individual refractive indices. (c) Coincidence probability, varying between 0 and about 0.7 depending on the refractive index $n_s$. The HOM dip appears around $n_s = 1.31$, where the transmittance is equal to the reflectance as shown in (b). Here, we assume that the wavelength of single photons is 800 nm.

In Fig. 3(a), for $\theta_{in} = 70°$ and $\lambda = 800$ nm, we compare the Fisher information of the HOM sensor $I^{(HOM)}$ with the classical benchmark $I^{(C)}$ that is maximized for an optimal value of $\phi_{\alpha\beta} = \pi/2$. It can be clearly seen that the HOM sensor outperforms the classical benchmark near $n_s = 1.32$ when the coincidence probability curve is steep [see Fig. 2(c)], yielding large derivative values in Eq. (18). When the HOM is maximized, i.e., $T = R$ near $n_s = 1.31$ referring to the spectra shown in Fig. 2(b), on the other hand, both the HOM and classical sensing schemes exhibit zero Fisher information. This implies infinite uncertainty in the estimation – a behavior that can be attributed to the small derivative value of the coincidence probability in Eq. (18), where the contribution of the other probabilities are minor. The value of the maximized Fisher information $I^{(HOM)}$ near $n_s = 1.31$ corresponds to the accuracy of the analyte's refractive index measurement, $\Delta n_{analyte} = 0.0154$, under the condition that saturates the Cramér-Rao inequality in Eq. (17). With an intensity of just a few picowatts, this has the potential to surpass the accuracy of the common technologies we aim to achieve [43].

To examine the relative quantum enhancement of the proposed HOM sensor, we define the ratio $G$ as

$$G = \frac{I^{(HOM)} - I^{(C)}}{I^{(C)}}. \qquad (19)$$

Values of $G > 0$ indicate the relative quantum enhancement, while values of $G \leq 0$ mean no quantum enhancement. This is elaborated in Fig. 3(b) for the Fisher information results shown in Fig. 3(a) for $\lambda = 800$ nm. The plot shows that the quantum enhancement reaches about 50 %, while ignoring the abrupt peak that occurs when both $I^{(HOM)}$ and $I^{(C)}$ approach zero near $n_s = 1.32$. The peak is due to the difference between $I^{(HOM)}$ and $I^{(C)}$ (the numerator in Eq. (19)) being larger than the value of $I^{(C)}$ (the denominator in Eq. (19)) as both $I^{(HOM)}$ and $I^{(C)}$ approach zero. In this case neither the quantum or classical case is favourable for sensing and thus the peak can be removed from the discussion of the enhancement. We also investigate the effect of the wavelength from 790 nm to 810 nm in Fig. 3(c), showing a wide region with 50 % quantum enhancement.

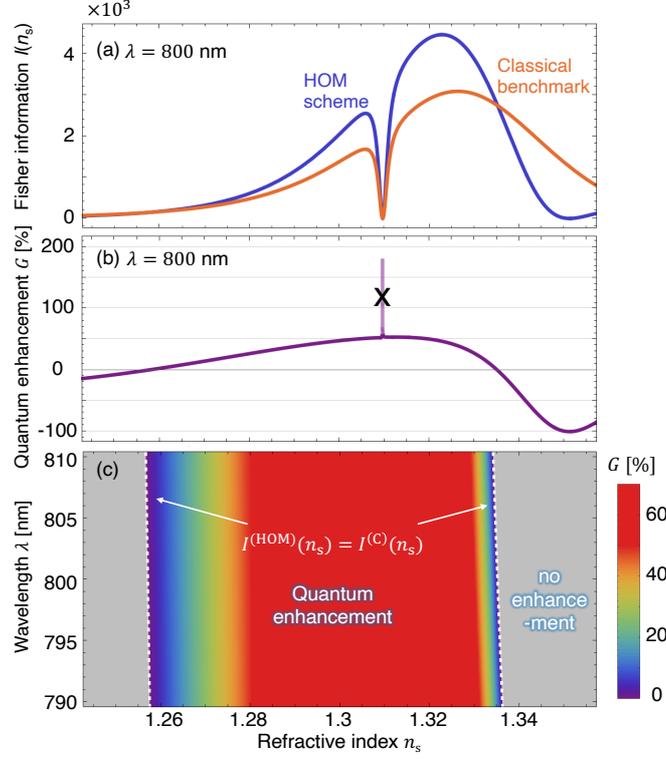

Fig. 3. (a) Fisher information for the quantum (HOM-based scheme) and coherent state input to the plasmonic BS. (b) Enhancement $G$. The peak near $n_s = 1.31$ abnormally occurs at a position where both Fisher informations go to zero. Thus, the peak is not useful. (c) The enhancement depends on refractive index $n_s$ and wavelength of the input photon. The colored area $G$ is larger than zero, meaning that HOM-based sensing has more information than the coherent state case. White dashed lines are contours for $G = 0$.

To understand the quantum enhancement shown in Fig. 3, we further decompose the Fisher information $I(n_s)$ of Eq. (18) according to the chain rule. It can be re-written as

$$I(n_s) = \sum_m \frac{1}{P(m|n_s)} \left( \sum_\tau \frac{\partial P(m|n_s)}{\partial \tau} \frac{\partial \tau}{\partial n_s} \right)^2 = \sum_{\tau,\rho} I_{\tau,\rho}(n_s|\{\tau,\rho\}) \frac{\partial \tau}{\partial n_s} \frac{\partial \rho}{\partial n_s}, \qquad (20)$$

where $\tau, \rho \in \{T, R, \phi_{\text{tr}}\}$ are dependent on the value of $n_s$. The Fisher information $I_{\tau,\rho}(n_s|\{\tau,\rho\})$ is defined as $\sum_m \partial_\tau P(m|n_s) \partial_\rho P(m|n_s)/P(m|n_s)$, which is the same as the definition of a typical Fisher information matrix [44]. Note that the derivative part on the right side, as simply calculated from classical wave theory (Eqs. (A4)- (A8)), is determined by the characteristics of the plasmonic BS. This calculation is irrelevant to the type of input state, thus yielding the same values for both the HOM and classical sensors. On the other hand, the Fisher information part $I_{\tau,\rho}$ plays a crucial role in distinguishing quantum enhancement [14, 45]. The terms associated with $\phi_{tr}$ can be shown to have a minor contribution to the quantum enhancement, so we focus on the role of the transmittance and reflectance, i.e., $I_{TT}$, $I_{TR}(= I_{RT})$, and $I_{RR}$.

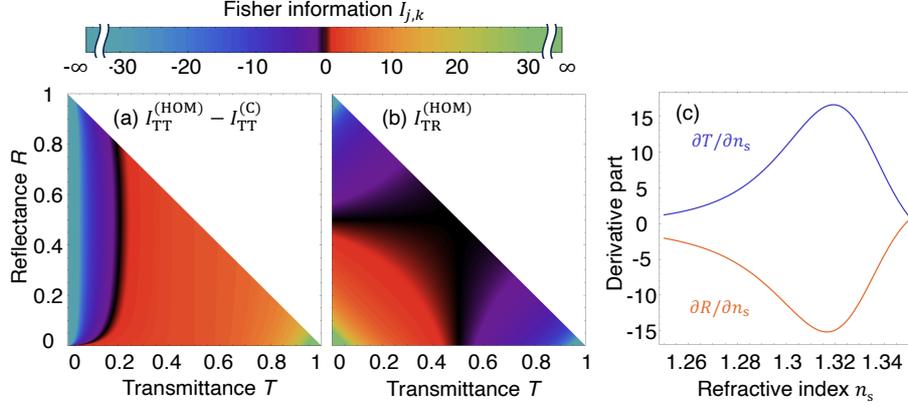

Fig. 4. Elements of Fisher information and derivative parts of the transmittance and reflectance. (a) Fisher information difference $I_{TT}^{(HOM)} - I_{TT}^{(C)}$. (b) Off-diagonal element $I_{TR}^{(HOM)}$. The elements (a) and (b) are plotted under assumption that $\phi_{tr} = \pi/2$. (c) Derivative parts of the transmittance and reflectance.

In Eq. (20), it is apparent that both $I_{TT}$ and $I_{RR}$, which are always positive according to statistical theory [44] and the associated derivative parts are all positive because they become squared if $\tau = \rho$ in Eq. (20). So, the quantum enhancement would be obtained when $I_{TT/RR}^{(HOM)} > I_{TT/RR}^{(C)}$ in a qualitative sense. Fig. 4(a) clearly shows the region with $I_{TT}^{(HOM)} > I_{TT}^{(C)}$. On the other hand, $I_{TR}$ can be positive or negative, as shown in Fig. 4(b). From Fig. 4(c), it is clear that the associated derivative part $(\partial_{n_s} T)(\partial_{n_s} R)$ is mostly negative, so a quantum enhancement would be achieved when $I_{TR} < 0$, which is shown in Fig. 4(b). This can be compared with the classical case for which $I_{TR}^{(C)} = 0$ due to the absence of a correlation between the two output ports (see Appendix B). Therefore, the interplay between the region with $I_{TT}^{(HOM)} > I_{TT}^{(C)}$ and $I_{TR} < 0$ approximately explains the quantum-enhanced region shown in Fig. 3. Note that the above behavior depends on the value of $\phi_{tr}$, but we assume $\phi_{tr} = \pi/2$ for simplicity. The actual values of $\phi_{tr}$ slightly deviate when the refractive index $n_s$ varies from 1.25 to 1.34, which is shown in Appendix B. Other sources of uncertainty, such as the incidence angle, polarization, prism refractive index, and film thickness, are described in an uncertainty budget in Appendix C. These uncertainties affect the precision of the HOM sensor to an extent ranging from $10^{-5}$ to $10^{-9}$. This is negligible compared to the lower bound of the uncertainty in Eq. (17) in the photon-starving regime, i.e. where it makes sense to use quantum measurement for maximal accuracy.

A final interesting observation is that in a previous report [19], it was shown that it is possible to mimic a HOM curve with the classical input we have considered here, by randomizing the phase $\phi_{\alpha\beta}$ as $-\pi/2$ and $\pi/2$. However, the Fisher information of the classical input state is maximized at both $\phi_{\alpha\beta} = -\pi/2$ and $\pi/2$, as shown in Fig. B3 in Appendix B. Thus, it is not possible to improve the Fisher information even by using a classical input state with randomized phase that mimics the HOM curve.

## 4. Conclusion

We have proposed and studied a quantum plasmonic sensing scheme exploiting a HOM interference through a plasmonic BS composed of a dual-Kretschmann configuration. The refractive index of an analyte embedded in the plasmonic BS can be estimated from a coincidence detection, and its sensing performance evaluated through the Fisher information within the framework of quantum parameter estimation. By comparing the plasmonic HOM sensor we propose with the classical benchmark that uses coherent states of light in the same plasmonic BS setup, we found that about a 50 % quantum enhancement is achieved over a wide range of wavelengths and refractive indexes even in the presence of Ohmic loss. The Ohmic loss could be improved by using longer wavelength photons, such as those in the C-band, where metals have lower loss. We expect this work to unlock potential advancements in quantum-enhanced sensing for various nanophotonic BS structures. In addition, this technique can be applied to sensors requiring the detection of small changes, such as gas or liquid detection and biomolecule sensing. The latter application often carries restrictions in the light exposure to avoid phototoxicity. Therefore, achieving maximal accuracy with a limited amount of resources (here, photons) requires the use of quantum measurement methods that can surpass classical sensors due to a quantum advantage [39, 46, 47].

## Acknowledgments

CL was supported by the National Research Council of Science & Technology (NST) Grant (No. CAP22052-000), Institute of Information & communications Technology Planning & Evaluation (IITP) Grant (No. 2022-0-00198), and Creation of the Quantum Information Science R&D Ecosystem (No. 2022M3H3A106307411) through the National Research Foundation of Korea (NRF), funded by the Korea government (MSIT). MST thanks the Department of Science and Innovation (DSI) through the South African Quantum Technology Initiative (SA QuTI), the National Research Foundation (NRF), and the Council for Scientific and Industrial Research (CSIR). JWY was supported by the Leader Researcher Program (NRF-2019R1A3B2068083).

## Appendix

### A. Transfer matrix

In this section, we provide the calculation of the transfer matrix in detail. The transfer matrix at a boundary is described as

$$M_{ab}^{b} = \frac{1}{t_{ba}} \begin{pmatrix} t_{ab}t_{ba} - r_{ab}r_{ba} & r_{ab} \\ -r_{ba} & 1 \end{pmatrix}, \quad (A1)$$

where $t_{ab(ba)}$ and $r_{ab(ba)}$ represent transmission and reflection coefficients from layer a to b (from b to a), respectively. A surface plasmon polariton is excited only by light in a transverse mode (parallel to the plane of incidence). So, the coefficients are written as

$$r_{ab} = \frac{n_a \cos(\theta_b) - n_b \cos(\theta_a)}{n_a \cos(\theta_b) + n_b \cos(\theta_a)} \quad (A2)$$

$$t_{ab} = \frac{n_a \cos(\theta_a)}{n_a \cos(\theta_b) + n_b \cos(\theta_a)}, \quad (A3)$$

where $n_{a(b)}$ denotes the refractive index of layer a(b). The angle $\theta_{a(b)}$ is the incident angle which satisfies Snell's law, $n_a \sin(\theta_a) = n_b \sin(\theta_b)$. We also have that $t_{ab}t_{ba} - r_{ab}r_{ba}$ is equal to 1. If the indices a and b are switched with each other, then $t_{ab} = t_{ba}$ and $r_{ab} = -r_{ba}$ are obtained from

Eq. (A2) and (A3). Thus, the transfer matrix is simplified as

$$M_{ab}^{b} = \frac{1}{t_{ab}} \begin{pmatrix} 1 & r_{ab} \\ r_{ab} & 1 \end{pmatrix}. \quad (A4)$$

Propagation through layer a is given by

$$M_{a}^{p} = \begin{pmatrix} e^{-i2\pi nd/\lambda} & 0 \\ 0 & e^{i2\pi nd/\lambda} \end{pmatrix}, \quad (A5)$$

where $n$ and $d$ denote refractive index and thickness of layer a, respectively. Here, $\lambda$ is the wavelength of the incident light. A transfer matrix for an $N$-layered structure is defined by combining Eqs. (A4) and (A5) as

$$M^{\text{total}} = M_{0,1}^{b} M_{1}^{p} M_{1,2}^{b} M_{2}^{p} \ldots M_{N-1,N}^{b} M_{N}^{p}. \quad (A6)$$

The transmission $t$ and reflection $r$ coefficients of the multi-layered structure are described as

$$t = 1/M_{11}^{\text{total}} \quad (A7)$$
$$r = M_{21}^{\text{total}}/M_{11}^{\text{total}}, \quad (A8)$$

where $M_{i,j}^{\text{total}}$ denotes an element of the matrix $M^{\text{total}}$. The refractive index of the sensing layer is encoded in the Fresnel coefficients at Eqs. (A4) and (A5) as a parameter. We considered a five-layered structure, as shown in Fig. 1: two glass prisms ($n$=1.5), two gold films, and one sample layer, where the refractive index of the gold film was taken from the work of Johnson and Christy [48].

### B. Fisher information

The phase $\phi_{\text{tr}}$ of our plasmonic beamsplitter can change because it has intrinsic loss. In Fig. B2, the values of $\phi_{\text{tr}}$ slightly changes when the refractive index $n_s$ varies from 1.25 to 1.34, which is the quantum enhancement range in Fig. 3

Here we provide the Fisher information of transmittance, reflectance, and phase which are not shown in main text. The diagonal terms, $I_{\text{TT/RR}}^{(\text{HOM}) \text{ or } (C)}$, are shown in Fig. B1. Note that $I_{\text{TT}}^{(\text{HOM}) \text{ or } (C)}$ in Fig. B1 (a, d) are symmetric with $I_{\text{RR}}^{(\text{HOM}) \text{ or } (C)}$ in Fig. B1 (b, e). Interestingly, the off-diagonal elements of the coherent state are zero in the entire region in Fig. B1 (c). $I_{\text{TR}}^{(C)}$ is zero because there is no correlation between the two ports.

In Fig. B3, the Fisher information varies with the phase $\phi_{\alpha\beta}$. The Fisher information is maximized at $\phi_{\alpha\beta} = -\frac{\pi}{2}$ and $\frac{\pi}{2}$. Thus, we choose $\phi_{\alpha\beta} = \pi/2$.

### C. Uncertainty budget

Here we provide an uncertainty budget for uncertainties, which can potentially influence estimating the analyte's refractive index with the HOM sensor. We assume that the Cramér-Rao inequality is saturated and thereby the uncertainties in Table C1 are obtained from $1/\sqrt{I(n_{\text{analyte}})}$. The relevant quantities in this study are input polarization, incidence angle, refractive index of the prisms, and thickness of the gold film. For the polarization, we assume that TM polarization is incident on the structure to excite a surface plasmon polariton, but in reality the input polarization could be slightly different from a TM mode because of the equipment's precision, such as a rotation stage. The angle $\gamma$ is defined as an angle separation from TM polarization ($\gamma = 0$). Also, the incidence angle can be affected by a rotation stage [49]. The effect of the refractive index of the prisms on uncertainty is provided by an estimate of the refractive index of BK7 [43]. It is possible that the thickness of the gold film has a ≤ 1 nm roughness, depending on the coating method [50]. The standard uncertainty is described as

$$\sigma_i = \frac{c_i \, s_i}{d_i} \quad (C1)$$

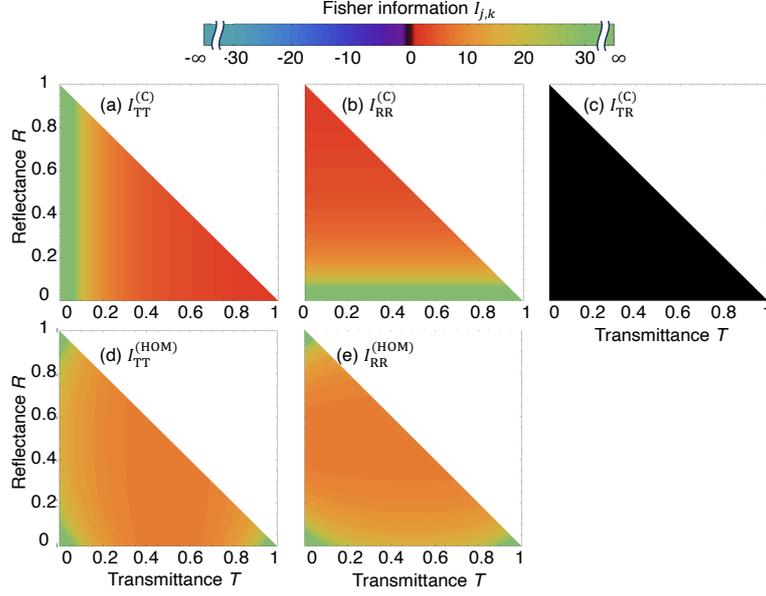

Fig. B1. Elements of the Fisher information for transmittance, reflectance, and phase. (a-c) Fisher information of the coherent state. The information $I^{(C)}_{TT/RR}$ are diagonal terms. The off-diagonal term of the coherent state, $I^{(C)}_{TR}$, is zero. (d and e) Fisher information of HOM interferometry. The elements of the Fisher information in (a)-(e) are shown under assumption that $\phi_{tr} = \pi/2$.

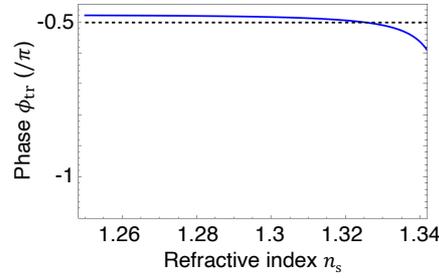

Fig. B2. Phase difference $\phi_{tr}$ curve between transmission and reflection. The dashed line is $-\frac{\pi}{2}$.

, where $\sigma_i$ is the standard uncertainty of $i$th source of uncertainty. $c_i$, $s_i$, and $d_i$ are a sensitivity, an uncertainty, and a divisor, respectively. A divisor describes a factor that converts a type of uncertainty in different probability distributions into its standard uncertainty. In this case, the divisor is unity since the sources of uncertainties have the normal distribution [51, 52].

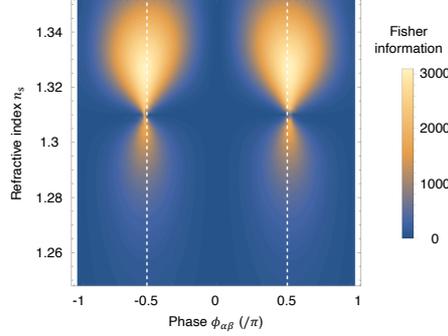

Fig. B3. Fisher information of the classical counterpart depending on phase $\phi_{\alpha\beta}$ and refractive index . The dashed line is $-\frac{\pi}{2}$ and $\frac{\pi}{2}$.

| Source of uncertainty | Uncertainty $s_i$ | Sensitivity $c_i$ | Standard uncertainty $\sigma_i$ |
|---|---|---|---|
| Input polarization $\gamma$ | 0.03 (°) | $3.00 \times 10^{-5}$ (1/°) | $9.00 \times 10^{-7}$ |
| Incidence angle | 0.03 (°) | $7.27 \times 10^{-4}$ (1/°) | $2.18 \times 10^{-5}$ |
| Refractive index of prism | $6.5 \times 10^{-5}$ (RIU) | $9.82 \times 10^{-2}$ (1/RIU) | $6.38 \times 10^{-6}$ |
| Thickness of gold film | $1 \times 10^{-9}$ (m) | 1.02 (1/m) | $1.02 \times 10^{-9}$ |

Table C1. Uncertainty budget. The unit of the standard uncertainty is refractive index unit. This table is considered at the case of the analyte's refractive index $n_{\text{analyte}} = 1.32$.

## D. Continuous mode

Here we provide the calculation of the Fisher information for the continuous mode case. For the continuous mode, the average of the number operator in Eqs. (13)–(15) can be rewritten as

$$\langle \hat{N}_1 \rangle = \int_0^\infty d\omega \, d\omega' \{|\psi(\omega,\omega')|^2 |t(\omega)|^2 + |\psi(\omega',\omega)|^2 |r(\omega)|^2\}, \tag{D1}$$

$$\langle \hat{N}_{1(2)}(\hat{N}_{1(2)} - 1) \rangle = \int_0^\infty d\omega \, d\omega' \{|\psi(\omega',\omega)|^2 |t(\omega')|^2 |r(\omega)|^2 + |\psi(\omega,\omega')|^2 |t(\omega)|^2 |r(\omega')|^2$$
$$+ \psi(\omega',\omega)^* \psi(\omega,\omega') t(\omega')^* t(\omega) r(\omega)^* r(\omega')$$
$$+ \psi(\omega',\omega) \psi(\omega,\omega')^* t(\omega') t(\omega)^* r(\omega) r(\omega')^*\}, \tag{D2}$$

$$\langle \hat{N}_1 \hat{N}_2 \rangle = \int_0^\infty d\omega \, d\omega' \{|\psi(\omega,\omega')|^2 |t(\omega')|^2 |t(\omega)|^2 + |\psi(\omega',\omega)|^2 |r(\omega)|^2 |r(\omega')|^2$$
$$+ \psi(\omega,\omega')^* \psi(\omega',\omega) t(\omega')^* t(\omega)^* r(\omega) r(\omega')$$
$$+ \psi(\omega',\omega') \psi(\omega,\omega')^* t(\omega') t(\omega) r(\omega)^* r(\omega')^*\}, \tag{D3}$$

where $\omega$ and $\omega'$ are the optical frequencies for the output ports. These averages are assigned to Eqs. (9)-(12). The Fisher information for the continuous mode case can be formulated using Eqs. (5)-(7) and (18). The average of the number operator of classical counterparts for the

continuous mode can be denoted as

$$\left\langle \hat{N}_1^{(C)} \right\rangle = \int_0^\infty d\omega\, d\omega' |t(\omega)\alpha(\omega)\xi(\omega) + r(\omega')\beta(\omega')\xi(\omega')|^2, \tag{D4}$$

$$\left\langle \hat{N}_2^{(C)} \right\rangle = \int_0^\infty d\omega\, d\omega' |t(\omega)\beta(\omega)\xi(\omega) + r(\omega')\alpha(\omega')\xi(\omega')|^2, \tag{D5}$$

where $\left\langle \hat{N}_{1(2)}^{(C)} \right\rangle$ is the average of the photon number at the output port of $\alpha$ (or $\beta$). We assume that $\int_0^\infty d\omega |\alpha(\omega)\xi(\omega)|^2 = 1 = \int_0^\infty d\omega |\beta(\omega)\xi(\omega)|^2$ for a fair comparison. The Fisher information of the classical counterparts for the continuous mode case can be formulated using Eq. (16) and (18). We assume that the frequency mode is Gaussian as

$$\xi(\omega) = e^{-2\ln 2 \left(\frac{\omega - \omega_0}{\Delta \omega}\right)^2}, \tag{D6}$$

$$\psi(\omega, \omega') = \xi(\omega)\xi(\omega'), \tag{D7}$$

where $\omega_0$ represents the central frequencies of the two photons. $\Delta\omega$ is the spectral bandwidths. The relative difference $D$ is defined as

$$D(n_s) = \left| \frac{I(n_s) - I(n_s)^{\text{cont}}}{I(n_s)} \right|, \tag{D8}$$

where $I(n_s)$ is for the single mode approximation treated in main text and $I(n_s)^{\text{cont}}$ denotes the Fisher information of the continuous mode case. The Fisher information for the HOM case is more robust to changes in the bandwidth of the photons than that the coherent state case in terms of the relative difference $D$, as can be seen in Fig. D1.

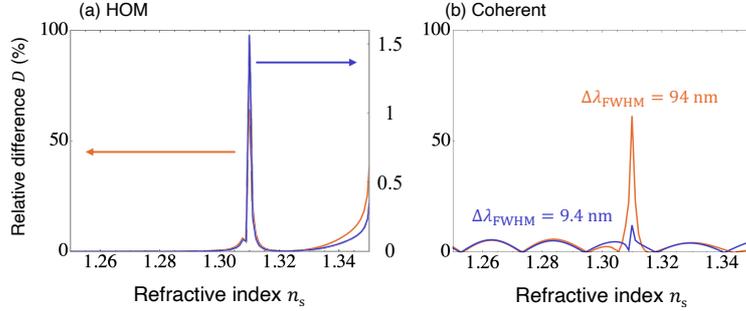

Fig. D1. Relative difference, $D$, for two cases using the continuous mode description: (a) HOM and (b) coherent state statistics. The blue line is for $\Delta\lambda_{\text{FWHM}} = 9.4$ nm. The orange represents the photons having a broad spectrum of 94 nm. The continuous modes of HOM are more robust to changes in the bandwidth of the photons than the coherent state.